\documentclass[preprintnumbers,aps,twocolumn,prd]{revtex4}
\usepackage{graphicx}
\usepackage{latexsym}
\usepackage{color}

\newcommand{\be}{\begin{equation}}
\newcommand{\ba}{\begin{eqnarray}}
\newcommand{\ee}{\end{equation}}
\newcommand{\ea}{\end{eqnarray}}

\newcommand{\barr}{\begin{array}}
\newcommand{\ear}{\end{array}}

\newlength\lfig
\begin{document}
\preprint{}

\vspace{2cm}

\title{Moduli evolution in the presence of thermal corrections}

\author{Tiago Barreiro}
\email{tiagobarreiro@fisica.ist.utl.pt}
\affiliation{IPFN, Instituto Superior T\'ecnico,
UTL, Av. Rovisco Pais 1, 1049-001 Lisboa, 
Portugal \\
Dept. de Matem\'atica, ULHT, Av. Campo Grande, 376, 1749-024 Lisboa, Portugal}
\author{Beatriz de Carlos}
\email{B.de-Carlos@soton.ac.uk}
\affiliation{ School of Physics \& Astronomy, University of Southampton, 
Highfield, Southampton SO17 1BJ, UK}
\author{Edmund J. Copeland}
\email{ed.copeland@nottingham.ac.uk}
\affiliation{The School of Physics and Astronomy, University of Nottingham, University Park, Nottingham NG7 2RD, UK}
\author{Nelson J. Nunes}
\email{nunes@damtp.cam.ac.uk}
\affiliation{DAMTP, Wilberforce Road, Cambridge CB3 0WA, UK}
\date{\today}
\begin{abstract}
We study the effect of thermal corrections on the evolution of moduli in effective Supergravity models. This is motivated by  previous results in the literature suggesting that these corrections could alter and, even, erase the presence of a minimum in the zero temperature potential, something that would have disastrous consequences in these particular models. We show that, in a representative sample of flux compactification constructions, this need not be the case, although we find that the inclusion of thermal corrections can dramatically decrease the region of initial conditions for which the moduli are stabilised. 
Moreover, the bounds on the reheating temperature coming from demanding that the full, finite temperature potential, has a minimum can be considerably relaxed given the slow pace at which the evolution proceeds.
\end{abstract}

\maketitle

\section{INTRODUCTION}

The study of moduli evolution is an active field of research in the context of the phenomenology and cosmology of string models. At the level of the $D=4$ effective theory, which we normally assume to be ${\cal N}=1$ Supergravity (SUGRA), moduli are complex scalar fields, some of which physically parametrise the size and shape of the six or seven original string dimensions that have been compactified. It is therefore mandatory that any realistic model provides, at the end of the day, moduli with a non trivial vacuum expectation value at the right scale (which is the Planck mass, $M_{\rm P}$ for conventional small extra dimensions), and that the minimum corresponds to almost Minkowski space and Supersymmetry broken, in order to connect with the Standard Model. Throughout the past twenty years there has been steady progress in our understanding of the dynamics of moduli, with two outstanding problems having to be addressed: the first one is the recurrence of a negative (i.e.  anti de Sitter, AdS) vacuum energy for every model in which moduli were successfully stabilised, while the second one is the fact that these potentials are so steep along certain directions that, from the dynamical point of view, it looked impossible that given any initial conditions away from their minimum, the moduli would end up in it. This issue was first pointed out by  Brustein and Steinhardt~\cite{Brustein:1992nk} and is commonly known as the problem of the "runaway dilaton". 

Concerning the first problem, namely the recurrence of AdS solutions within all successful attempts at stabilising moduli, recent developments in the context of flux compactifications in type IIB string theory~\cite{Giddings:2001yu}  have opened up new ways of trying to achieve either Minkowski or de Sitter (dS) vacua. In particular,  the mechanism presented by Kachru et al.~\cite{Kachru:2003aw} (KKLT from now on) realised this by adding D-terms to a SUSY-preserving, AdS F-term vacuum. Although not entirely correct in the context of Supergravity (see~\cite{Choi:2005ge,deAlwis:2005tf} for some criticism, and~\cite{Burgess:2003ic,Achucarro:2006zf} for proposed solutions), its main features have triggered an enormous amount of interesting work, and subsequent progress in this field through the past years (in particular, for explicit string realisations, see~\cite{Haack:2006cy,Cremades:2007ig}). 

Also on the topic of runaway moduli, in general, substantial developments have taken place in the past decade. For a range of initial conditions the problem can be alleviated considerably by considering a background perfect fluid which decelerates the fields and prevents them from passing the barrier dividing the physical vacuum from the runaway one~\cite{Kaloper:1991mq,Barreiro:1998aj,Huey:2000jx,Barreiro:2000pf,Brustein:2004jp,Kaloper:2004yj,Barreiro:2005ua}. Thermal corrections, however, are potentially dangerous as they modify the shape of the potential and, at high temperatures, the physical vacuum is entirely lost~\cite{Buchmuller:2004xr,Buchmuller:2004tz}. However, there is a  potential limit to the extent to which this argument can be used. In the very early Universe as the value of the Hubble 
parameter, becomes close to the Planck scale, scattering processes are unable to establish thermal 
equilibrium because they do not have sufficient time compared to the expansion rate of the Universe \cite{Ellis:1979nq}.  In this era thermal corrections arising out of these scatterings would not be present, and so the physical vacuum would not be destabilised. In the context of the Standard Model
Enqvist and Sirkka  considered the thermalisation of a hot QCD gas in the early universe, and calculated the critical temperature above which the Universe can not thermalise to be $T_{\rm crit} = 3 \times 10^{14}$ GeV \cite{Enqvist:1993fm}.  In Ref.~\cite{Lalak:2005hr} the authors used this argument in considering racetrack inflation and assisted moduli stabilisation.

In this work we present a detailed quantitative analysis of the effect of these thermal corrections on the region of initial conditions leading to stabilisation of the moduli, for a given SUGRA model.  Given that we are working in regimes beyond the standard model, where we do not have a proper handle on the conditions under which we will move out of thermal equilibrium, we adopt two approaches. The more theoretical is to accept that the Universe may have been in a period of thermal equilibrium close to the Planck scale and investigate the impact of thermal corrections in those regimes. The second is to take seriously the QCD bound, and investigate the impact of the corrections in these lower temperature regimes.
In both cases we will conclude that thermal corrections do decrease the area of the region of stabilisation by a similar relative amount. 
This effect from the thermal coupling will act on top of the previous results, where the absolute size of the stabilization region decreases for a smaller initial density of the background fluid $\rho_r^{\rm init}$ \cite{Barreiro:2005ua}.

%%%%%%%%%%%%%%%% SET UP  %%%%%%%%%%%%%%%%%%%%%%%%%%%%%%
\section{EQUATIONS OF MOTION} 
\label{sec:setup}

In this work, we will be studying two string theory models that can be described by a four dimensional ${\cal N}=1$ effective Supergravity theory with action of the form
\begin{equation}
S=-\int d^4x  \sqrt{-g} \left(\frac{1}{16 \pi G}R- {\cal L}_\Phi + F(g,T) \right)  \;\;,
\label{eq:action}
\end{equation}
where
\be
{\cal L}_\Phi = - K_{i\bar{\jmath}}\partial_{\mu}\Phi ^{i}\partial^{\mu}{\bar{\Phi}^{\bar{\jmath}}}-V \,,
\ee
and  $K_{i\bar{\jmath}}= \partial^{2}K/\partial\Phi^{i}\partial\bar\Phi^{\bar{\jmath}}$ is the K\"ahler metric; $\Phi^{i}$ are complex moduli scalar fields; $V(\Phi)$ is the scalar potential and $G$ is the 4-dimensional Newton constant. The free energy $F(g,T)$ acts as a Lagrangian density of matter fields. For a ${\rm SU}(N_c)$ gauge theory with $N_f$ multiplets at high temperature $T$, the free energy has a perturbative expansion in terms of the gauge coupling $g = g(\Phi_R)$, where $\Phi_R$ denotes the real part of the moduli fields $\Phi$,
\be
F(g,T) =  \left(a_0 + a_2 \, g^2 \right) T^4 \,,
\ee
and the parameters $a_0$ and $a_2$ are given by
\ba
\label{a0}
a_0 &=& -\frac{\pi^2}{24} \left(N_c^2 + 2 N_c N_f-1\right) \,, \\
\label{a2}
a_2 &=& \frac{1}{64} (N_c^2-1) (N_c+3N_f) \,.
\ea
We will be treating $a_0$ and $a_2$ as variables which can be varied in order to test the results for a range of possible values of $N_c$ and $N_f$.
We can obtain the energy density and pressure of this thermal fluid as $p_r = -F$ and $\rho_r = -p_r + T d p_r/d T$, hence
\be
\label{rhor}
%\rho_r = -3 (a_0 + a_2\, g^2) T^4 \,.
\rho_r = -3 \, a_0 (1 + r g^2) T^4 \, ,
\ee
where $r = a_2/a_0$.

In principle, a second component of relativistic particles that only interacts with the moduli fields gravitationally can also be present in the dynamics. 
Assuming both components of radiation to be in thermal equilibrium, the non interacting radiation will have an energy density given by
$\rho_{\rm B} = \pi^2 g_* T^4/30$ where $g_*$ is the number of effective degrees of freedom at the temperature $T$ of the thermal fluid $\rho_r$.
In this case, the equations of motion above are still valid with $a_0$ being replaced by $a_0 \rightarrow a_0 -\pi^2 g_*/90$. The effective $|a_0|$ can then be very large , so that $r \approx 0$, effectively washing out the effects of the thermal corrections on the moduli evolution.

The effective scalar potential for the moduli in four dimensional ${\cal N}=1$ Supergravity is given by 
\begin{equation}
V = e^{K}(K^{i\bar{\jmath}}D_iWD_{\bar{\jmath}}{\bar W}-3W\bar{W}) \;\;,
\label{eq:potential}
\end{equation}
where $K^{i\bar{\jmath}}$ is the inverse K\"ahler metric and
$D_iW=\partial_{i}W+\partial_i K \,W$ is the
 K\"ahler covariant derivative acting on the superpotential. In general, the K\"ahler potential $K$ is a function of the real parts of the fields and, for most string compactifications, acquires the form
 \be
 K = -\sum_i\ln(\Phi_i+\bar{\Phi}_i) \;\;,
 \label{kahler}
 \ee
where the sum is understood over all moduli $\Phi_i$. As for the superpotential, which encodes the dynamics of these fields, a combination of flux terms (normally polynomials in the different moduli) and non perturbative effects (instantons and  gaugino condensation being the most well-known ones \cite{Nilles:1982ik,Ferrara:1982qs,Derendinger:1985kk,Dine:1985rz}) will provide the potential with a non trivial vacuum structure. For the purposes of showing the effects of thermal corrections, we use the toy models of KKLT~\cite{Kachru:2003aw} and Kallosh and Linde~\cite{Kallosh:2004yh} (KL from now on), given that they capture the essential features of the cosmological problems usually attributed to moduli.

The equations of motion follow from the variation of the action (\ref{eq:action}). Considering homogeneous fields evolving in a spatially flat Friedmann-Robertson-Walker spacetime background, the equations of motion for the complex fields yield
\begin{eqnarray}
\ddot{\Phi}^{i}+3H\dot{\Phi}^i+\Gamma^{i}_{jk}\dot{\Phi}^{j}
\dot{\Phi}^{k}+K^{i\bar{\jmath}} \partial_{\bar{\jmath}} V =  \nonumber \\
%= -a_2  T^4 K^{i\bar{\jmath}} \partial_{\bar{\jmath}} g^2   
= \frac{r \, \rho_r}{3 (1+rg^2)} K^{i\bar{\jmath}} \partial_{\bar{\jmath}} g^2    \,,
\label{eq:fulleom}
\end{eqnarray}
where 
\mbox{$\dot{\Phi}^i=\partial{\Phi}^i/\partial{t}$},
\mbox{$\partial_{\bar{\jmath}}V=\partial{V}/\partial{\bar{\Phi}^{\bar{\jmath}}}$},
and the connection on the K\"ahler manifold has the form 
\begin{equation}
\Gamma^{i}_{jk}=K^{i \bar l}\frac{\partial K_{j\bar l}}{\partial \Phi^{k}} \; .
\label{eq:gamma}
\end{equation}
In addition, the Hubble rate $H \equiv \dot{a}/a$, where
$a(t)$ is the scale factor of the Universe, is given by the Friedmann equation
\begin{equation}
3H^2= M_{\rm P}^{-2} (\rho_{\Phi}+\rho_r) = M_{\rm P}^{-2}
(K_{i\bar{\jmath}} \dot{\Phi}^{i}\dot{\bar{\Phi}}^{\bar{\jmath}}+V+\rho_r) \;\;,
\label{eq:fullfriedman}
\end{equation}
with
$ M_{\rm P}^{-2} = 8\pi G$ (or $M_{\rm P} = 2 \times 10^{18}$ GeV) 
and $\rho_\Phi = K_{i \bar{\jmath}} \dot{\Phi}^i \, \dot{\bar{\Phi}}^{\bar{\jmath}} + V $ and $\rho_r$
are the energy densities of the evolving moduli fields and of the thermal fluid, respectively. In what follows we set $M_{\rm P} = 1$.

We need to understand now how the temperature $T$ relates to the values of the fields and the scale factor of the universe. To this end we note that 
the equations of motion for the scalar fields can be rewritten as
\be 
\dot{\rho}_\Phi = -3 H (\rho_\Phi + p_\Phi)  + \frac{1}{3} \partial_i \rho_r \dot{\Phi}^i +
 \frac{1}{3} \partial_{\bar{\jmath}} \rho_r \dot{\bar \Phi}^{\bar{\jmath}} \;,
 \ee
where the pressure of the moduli fields is defined as $p_\Phi = K_{i \bar{\jmath}} \dot{\Phi}^i   \, \dot{\bar \Phi}^{\bar{\jmath}} - V$. By requiring conservation of the total energy density $\rho = \rho_\Phi + \rho_r$ we must have
\be 
\dot{\rho}_r = -4 H \rho_r  - \frac{1}{3} \partial_i \rho_r \dot{\Phi}^i -
 \frac{1}{3} \partial_{\bar{\jmath}} \rho_r \dot{\bar \Phi}^{\bar{\jmath}} \,,
 \ee 
which, upon integration, gives a solution for $\rho_r$ of the form
\be
\rho_r = \rho_r^{\rm init} \left(\frac{a_{\rm init}}{a}\right)^4 \left(\frac{1+r g^2(\Phi_R^{\rm init})}{1+r g^2(\Phi_R)} \right)^{1/3} \,.
\ee
Comparing with Eq.~(\ref{rhor}), the evolution of the temperature can be seen to be
\be
T = T_{\rm init} \frac{a_{\rm init}}{a} \, \left(\frac{1+ r g^2(\Phi_R^{\rm init})}{1+r g^2(\Phi_R)} \right)^{1/3} \,.
\ee

It is worth splitting the equations
of motion for the complex scalar fields into those for their real and imaginary parts
\begin{eqnarray}
\label{eq:real}
\ddot{\Phi}^{i}_{R}+3H\dot{\Phi}^{i}_{R}+\Gamma^{i}_{jk}(\dot{\Phi}^{j}_R
\dot{\Phi}^{k}_R-\dot{\Phi}^{j}_I\dot{\Phi}^{k}_I)+\frac{1}{2}K^{i\bar{\jmath}}\partial_{j_R}V  \label{eq:reom} \\
= \frac{1}{6} K^{i\bar{\jmath}} \partial_{j_R} g^2  \frac{r \, \rho_r}{1+rg^2} \;\;,
\nonumber
\end{eqnarray}
\begin{equation}
\label{eq:imag}
\ddot{\Phi}^{i}_I+3H\dot{\Phi}^i_I+\Gamma^{i}_{jk}(\dot{\Phi}^{j}_I\\
\dot{\Phi}^{k}_R+\dot{\Phi}^{j}_R\dot{\Phi}^{k}_I)+\frac{1}{2}K^{i\bar{\jmath}}\partial_{j_I}V=0 \;\;,
\label{eq:ieom}
\end{equation}
 where now $\Phi^{i}_{R}$ ($\Phi^i_I$) refers to the real 
 (imaginary) part of the scalar fields and  $\partial_{j_R}$
($\partial_{j_I}$) are
used to denote the derivative of the potential with respect
to the real (imaginary) parts of the fields respectively.  
We note that, given an initial value of the energy density of the thermal fluid, $\rho_r^{\rm init}$, the equations of motion and the Friedmann equation only depend on the ratio $r = a_2/a_0$ but not on the specific values of $a_0$ and $a_2$.

Given Eqs.~(\ref{eq:fulleom}) and (\ref{eq:real}) we can define an effective scalar potential for the fields,
\begin{equation}
V_{\rm eff} (\Phi) = V(\Phi) - \frac{1}{3} \frac{r\, \rho_r \, g^2}{1+r \,g^2} \,,
\end{equation}
up to a constant. With a coupling constant of the form \cite{Buchmuller:2004tz}
\begin{equation}
g^2 = \frac{c}{\Phi} \,,
\end{equation}
where $c$ is a constant, it is clear that, at high temperature (large $\rho_r$), the effective potential can look very different from $V$ and, in particular, it can be devoid of a minimum, limiting the stabilisation of the moduli or the maximum temperature allowed. This is the problem raised in Ref.~\cite{Buchmuller:2004tz}. 

%%%%%%%%%%%%%%%%%%%%%%%%%%%%%%%%%%%%%%%%%%%%%%%%%%%%%%%%%%%%%%%%%%%%%%%%%%%
\section{KKLT model}
\label{sec:KKLT model}
The possibility of finding  de Sitter vacua in string theory  with a stabilized volume modulus, $\sigma$,  was put forward in Ref.~\cite{Kachru:2003aw}, and has been widely adopted in subsequent work. The key ingredient was to consider the combination of non perturbative effects and an additional flux term in the superpotential
\be
W = W_0 + A e^{-\alpha \sigma} \:\:,
\label{superpot}
\ee
which, together with the usual K\"ahler potential
\be 
K = -3 \ln (\sigma+\bar{\sigma}) \;\;,
\ee
defines the $F$-term of the SUGRA potential, see Eq.~(\ref{eq:potential}). It has been known for many
years now that, in this context, it is possible to stabilize $\sigma$, although giving rise to an AdS  vacuum. As pointed out in Ref.~\cite{Kachru:2003aw}, however, if  contributions from either anti-D3 or D7 branes are included, an additional D-type term of the form
\be
V_D = \frac{C}{\sigma_R^3} \;\;,
\label{Dpot}
\ee
is generated, where we write $\sigma=\sigma_R+i \sigma_I$.
By suitably tuning the value of $C$ one can move to a de Sitter -- or even Minkowski -- vacuum.
The scalar potential for $\sigma$ has an extremum in $\sigma_I$ for $\alpha \sigma_I= n \pi$, with $n$ an integer. Depending on the sign of $W_0 \, {\rm cos}(\alpha\sigma_I)$ this can be either a maximum or a minimum. 

\begin{figure}[!ht]
\includegraphics[width=\lfig]{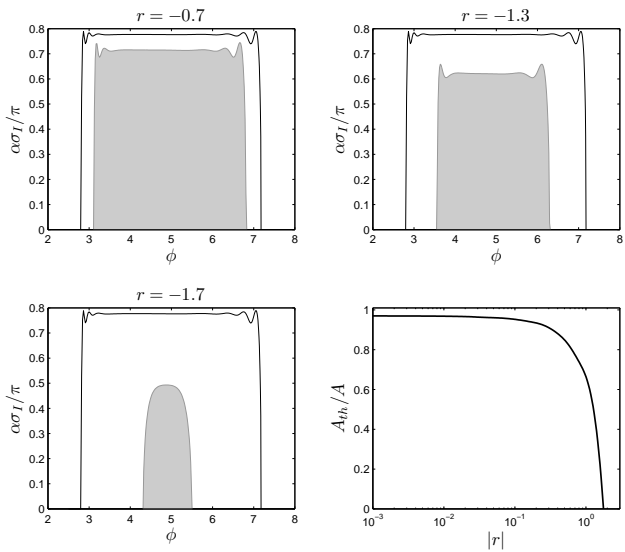}
\caption{\label{boundsa} The panels with the shaded regions show the regions of initial conditions ($\phi$, $\sigma_I$) -- in Planck units -- that lead to stabilisation of $\sigma$ at the minimum of the potential for the KKLT model in the presence of thermal corrections. The solid black line corresponds to the region of stabilisation in the presence of a perfect fluid {\bf (r=0)} with the same initial energy density $\rho_r^{\rm init}$. The lower right corner shows the ratio of the areas of these regions against the ratio $r = a_2/a_0$. We have set $\rho_r^{\rm init} = 10^{-4} M_{\rm P}^4$.}
\end{figure}

\begin{figure}[!ht]
\includegraphics[width=\lfig]{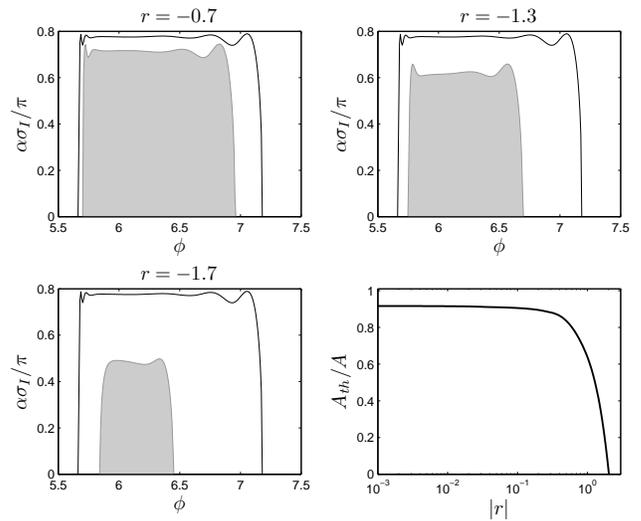}
\caption{\label{boundsa2} Same as Fig.~\ref{boundsa} but setting $\rho_r^{\rm init} = 10^{-13} M_{\rm P}^4$.}
\end{figure}

In this work we are interested in studying the cosmological evolution of
the field $\sigma$ as it rolls towards its minimum. Previous analysis addressing the same 
issue were published in Refs~\cite{Brustein:2004jp,Barreiro:2005ua}, however, without taking into account the effect of the thermal corrections. Here, we compare the profile and area of the region of initial positions of the field, $\sigma_{\rm init}$ that leads to stabilisation when the field rolls in the presence of a background perfect fluid with the same region of stabilisation when the field evolves in the presence of thermal corrections.
We illustrate that comparison for the KKLT model in Fig.~\ref{boundsa} and Fig.~\ref{boundsa2}  for two values of the initial energy density of the thermal fluid, $\rho_r^{\rm init}$. The first corresponds to the very high energy regime $\rho_r^{\rm init} = 10^{-4} M_{\rm P}^4$, where the second corresponds to a value which satisfies the condition for thermal equilibrium obtained in \cite{Enqvist:1993fm} for $|a_0| < {\cal O}(100)$, namely $\rho_r^{\rm init} = 10^{-13} M_{\rm P}^4$. In both cases the stabilisation region is shown with thermal corrections (shaded areas) for given values of the ratio $r = a_2/a_0$ against the stabilisation region with  just a perfect fluid of radiation (or equivalently, with $r = 0$) for the same initial $\rho_r$ (solid black line).

It proves convenient to work with the canonically normalized field $\phi = \sqrt{3/2} \ln \sigma_R$, instead of $\sigma_R$ itself.  In the lower right panel of the figures we draw the ratio between the area of the two stabilisation regions --- i.e. the one with thermal corrections ($A_{\rm th}$) and  the one without ($A$) --- against $|r|$. 

We note that, because the energy density of the thermal fluid must be positive definite,  and taking the effective 4D Yang-Mills coupling on the D7 brane $g^2 = 4\pi/\sigma_R$, we must ensure that
\begin{equation}
-\frac{1}{4\pi} e^{\sqrt{2/3} \, \phi} < r \le 0 \;\;.
\end{equation}
This means that we need to impose a larger lower bound on the initial values of $\phi$ as we increase $|r|$. In practice this bound is only effective for large initial $\rho_{r}^{\rm init}$ since the stabilisation region decreases naturally for lower initial temperatures.

Returning to Figs.~\ref{boundsa}--\ref{boundsa2}, we show the regions of initial conditions that lead to the field stabilising in the minimum of its potential. We used the same values of the parameters for the model as in Ref.~\cite{Brustein:2004jp}, namely $A=1.0$, $\alpha=0.1$, $C=3 \times 10^{-26}$, and $W_0$ negative (with ${\rm cos}(\alpha \sigma_I)=1$) such that the minimum at $\sigma_I=0$ is  supersymmetric. 
Comparing the two figures, we can see that the stabilisation region for the lower initial temperature is considerably smaller, a result that was shown previously for the evolution without a thermal coupling. The effect from the thermal coupling can be more clearly seen in the lower right plots, where the ratio of the two stabilisation areas is shown against $|r|$. We see that in both cases increasing the strength of the thermal coupling (that is, the relative size of $a_2$) can effectively  eliminate the stabilisation region. On the other hand, the curves for the two different choices of initial $\rho_{r}^{\rm init}$ are very similar, meaning that the relative effects of the thermal coupling seem to be independent of the initial value of the background. In both cases, a value of $|r| \lesssim 1$ implies that the stabilisation region is decreased by less than $50\%$. 

Using Eq.~(\ref{rhor}) with $a_0 = -100$, we can see that our choices for the inital background energy density correspond to initial temperatures of $T \sim 10^{-2} M_{\rm P}$ and $T \sim 10^{-4} M_{\rm P}$. However the dilaton potential only develops a minimum at a temperature $T_{\rm crit} \sim 10^{-8} M_{\rm P}$, so that we can have stabilisation of the dilaton even when the evolution starts at a temperature where the minimum does not exist.
 
\begin{figure}[!ht]
\includegraphics[width=\lfig]{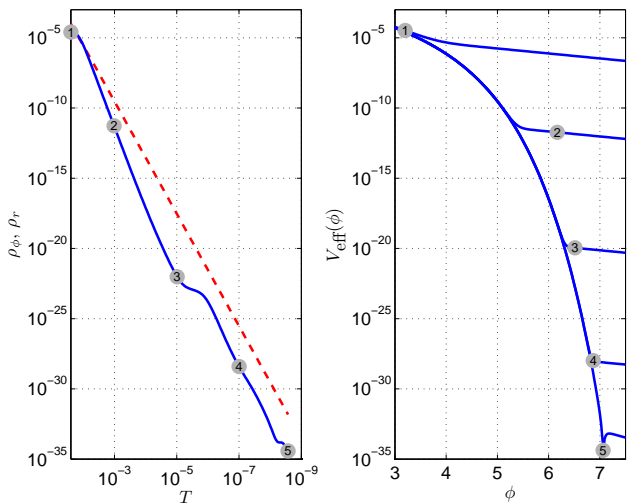}
\caption{\label{evrho} In the left panel we show the evolution of the energy densities of the field, $\rho_\phi$ (solid line) and of the thermal fluid, $\rho_r$ (dashed line) in the KKLT model, as a function of the temperature $T$ (all in Planck units) . In the right panel we show the profile of the effective potential at five different values of the temperature (solid lines) and the position of the field at these different stages (circles), as a function of $\phi$ (also in Planck units). We  have set $r = -0.2$, $a_0 = -100$, $\rho_r^{\rm init} = 10^{-4}$, $\phi_{\text{init}} = 3.2$ and $\sigma_{I} = 0$.}
\end{figure}

This behaviour can be easily understood. 
Thermal corrections act as a background fluid in that they bring an extra contribution into the Friedmann equation. 
Though at high temperatures the structure of the effective potential is spoiled (that is, we have no minimum), the extra contribution reinforces the frictional term in the equation of motion and forces the field to slow down its evolution as when in the presence of a non interacting perfect fluid. Therefore, for suitable  initial conditions, the field can approach the value corresponding to the minimum of the $T=0$ potential,
$\langle \phi \rangle_{T = 0}$, when the thermal corrections have already become negligible, i.e. the temperature has decreased below the critical temperature for which the minimum is created. In Fig.~\ref{evrho}, we show, in the left panel, the evolution of $\rho_{\phi}$ and $\rho_r$ as a function of the temperature for the model used in  Fig.~\ref{boundsa}, having singled out five values of the temperature. In the right panel we can see the profile of the effective potential for these five values of the temperature with the corresponding value of the field at that particular time. We can see, in position 3, at $T = 10^{-5} M_{\rm P}$, that the field is evolving in an effective potential without a minimum but that, in position 5, the effective potential has recovered the minimum  and the field has been trapped in it. 

We observe that, initially, the large kinetic energy  forces the field to evolve to the flatter region of the potential where the field effectively freezes due to the large frictional term in the equation of motion given that $\rho_r \gg \rho_{\phi}$ holds. As the universe expands the thermal corrections continue to decrease in magnitude and the field reenters the steep slope of the potential which admits a scaling-like regime of evolution i.e., the field restarts rolling with a kinetic energy that is proportional to the potential energy. It is this property that prevents the field from gaining kinetic energy and going over the barrier that separates the physical minimum from the runaway one.  In other words this is why the modulus
does not get driven to larger values more quickly. For an analytic description of this stabilisation mechanism due to scaling, see \cite{Barreiro:1998aj,Barreiro:2000pf,Brustein:2004jp,Barreiro:2005ua}.

%%%%%%%%%%%%%%%%%%SECOND MODEL%%%%%%%%%%%%%
\section{Kallosh-Linde model}
\label{sec:KLmodel}

\begin{figure}[]
\includegraphics[width=\lfig]{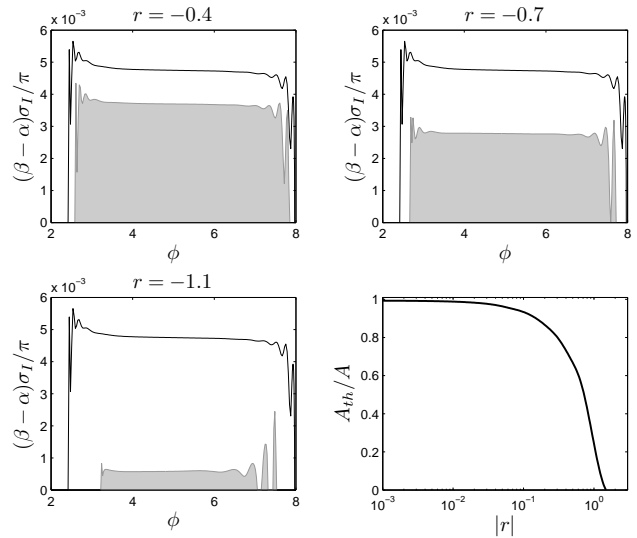}
\caption{\label{boundsb}  Same as in Fig.~\ref{boundsa} for the Kallosh-Linde model with $\rho_r^{\rm init} = 10^{-4} M_{\rm P}^4$.}
\end{figure}

\begin{figure}[]
\includegraphics[width=\lfig]{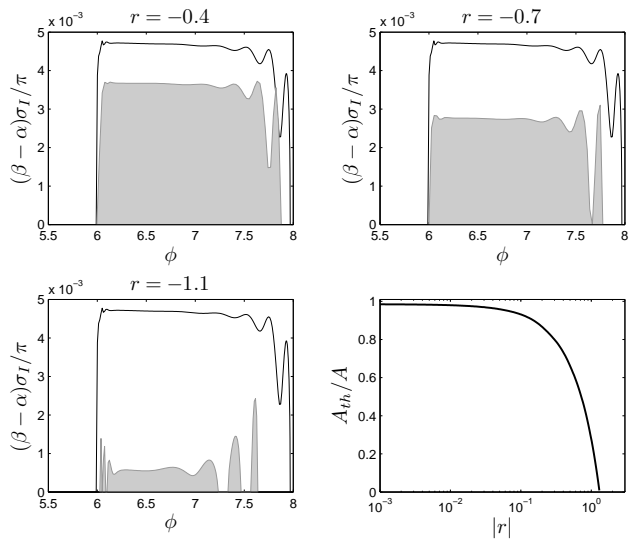}
\caption{\label{boundsb2}  Same as in Fig.~\ref{boundsa} for the Kallosh-Linde model with $\rho_r^{\rm init} = 10^{-13} M_{\rm P}^4$.}
\end{figure}

The Kallosh-Linde model (KL) \cite{Kallosh:2004yh} generalizes the original version of the KKLT model by admiting two components
in the superpotential 
\begin{equation}
W = W_0 + A e^{-\alpha \sigma} + B e^{-\beta \sigma} \,.
\label{WAB}
\end{equation}
A particular example was investigated in a previous publication~\cite{Barreiro:2005ua}. The parameters of the model were set to  $A = 1$, $B = -1.5$, $C = 0$, 
$\alpha = 2 \pi/100$, $\beta = 2\pi/99$ and $W_0$ such that there is a supersymmetric minimum with zero cosmological constant, i.e. $W_0$ is such that both $W$ and $F_{\sigma} \equiv K_{\sigma} W+W_{\sigma}$ vanish at some $\sigma_R=\sigma_{\rm crit}$, $\sigma_I=0$.  Furthermore,
there are a series of supersymmetric, AdS minima. 
\marginpar{$\star$}
In Figs.~\ref{boundsb}-\ref{boundsb2} we show how the stabilisation region with thermal corrections varies with $r$ and, for comparison, we plot it against the stabilisation region when a perfect fluid of radiation is present (or equivalently, when $r=0$). As for the KKLT model,
we see that for both energy regimes the profile of the stabilisation regions is practically unmodified for small values of $|r|$ and that their area decreases as $|r|$ becomes larger, eventually vanishing for sufficiently high values of $|r|$.
Furthermore, we see again that this effect of the thermal coupling on the stabilisation region is similar for the two initial values of $\rho^{\rm init}_r$ in relative terms, though a smaller initial value of the initial energy density in the background makes the stabilisation areas smaller in absolute terms. The thermal coupling is slightly more effective in reducing the stabilisation area in this KL model.

%

%%%%%%%%%%%%%%%%%%%%CONCLUSIONS%%%%%%%%%%%%%%%%%%%%%

\section{Discussion}
\label{sec:conclusions}

We have seen in the two models investigated that the inclusion of thermal corrections in  the moduli fields evolution can have a very similar effect to a perfect fluid of non interacting radiation, in the sense that their contribution  increases the friction, decelerating further the fields thus helping in their stabilisation at the desired minimum. 
We saw that the relative effect of the thermal coupling in the stabilisation area of the dilaton is nearly independent of the actual initial value of the background energy density. As we increase the value of $|r|$ (that is, we increase the relative value of $a_2$), the stabilisation region becomes smaller, and eventually disappears. Note however, that in both models, values of $|r| \lesssim 1$ will only decrease the area for at most $50 \%$. For reference, to have $|r| > 1$ we need  $N_c \geq 17$, whereas for $|r| > 1.5$ we need $N_c \geq 26$. The values of $|r|$ obtained strictly from the thermally coupled fields only gives us a lower bound on the stabilisation area. In principle, we will also have fields in the background without a thermal coupling to the dilaton in the background. In this case, we will have an effective value for $|r| \ll 1$, reducing considerably the global effect of the thermal coupling. Namely for $|r| < 10^{-2}$, the difference between the two scenarios is of less than $1 \%$.
On the other hand, it should also be clear that a larger 
value of the background energy density leads to a  wider region of stabilisation, as the field can enter a scaling regime earlier.

The existence of a minimum in the potential is usually seen as a necessary condition for the stabilisation of a moduli field, from which upper bounds to the reheating temperature can be obtained. We have seen that, if we allow for the field to evolve, the stabilisation is also dependent on its initial condition; and a minimum in the potential need not be present initially.
From Eq.~(\ref{rhor}) we get that for $|r| \lesssim 1$ the relation
between the initial value of the temperature and energy density is approximately 
\begin{equation}
T_{\rm init} \approx \left(\frac{\rho_r^{\rm init}}{-3 a_0}\right)^{1/4} \,,
\end{equation}
which means that, for values, $a_0 \approx - 100$ and $\rho_r^{\rm init} \approx 10^{-4} M_{\rm P}^4$ or $\rho_r^{\rm init} \approx 10^{-13} M_{\rm P}^4$, the initial value of the temperature can be as large as 
$10^{-2} M_{\rm P}$ or $10^{-4} M_{\rm P}$, respectively. These are above the usual upper limit given for stabilisation, $T< 10^{-8} M_{\rm P}$ suggested in Ref.~\cite{Buchmuller:2004tz} 
obtained by requiring the minimum to appear in the effective dilaton potential. 

Even though we do consider sectors different from the (MS)SM as the source
for the thermal bath, the simplest scenario would be to assume that the SM
fields would eventually enter thermal equilibrium at the same temperature,
assuming a smooth evolution of all the background fields. Of course this
does not need to be the case, and in that instance a different later
background with a higher temperature that interacts with the dilaton could,
in principle, destabilize the field again. Assuming the field to be already
in its minimum, this would have to be at a sufficiently high temperature to
effectively remove the minimum.

Though the mechanism here described seems very general,
it would be interesting to evaluate its robusteness for other scenarios such as the KKLT model coupled to a Polonyi field
studied in Ref.~\cite{van de Bruck:2007jw}. One other aspect is that to large temperatures correspond large thermal fluctuations which may give rise to large spatial inhomogeneities in the moduli. A quantitative investigation of these effects and their cosmological implications 
deserves a complete study.

%========================ACKNOWLEDGEMENTS=============================
\begin{acknowledgments}
TB is supported by FCT, under grant BPD/3512/2000. NJN is supported by STFC. We would like to thank Michael Ratz for early discussions on this project. We are grateful to Tony Riotto for his comments on the initial version of this paper and to the referees for their useful comments. 
\end{acknowledgments}

%=====================================================================
%=========================BIBLIOGRAPHY================================

\end{document}